\documentclass[aps,superscriptaddress,reprint,nofootinbib]{revtex4-1}

\usepackage{graphicx}
\usepackage{amsmath}
\usepackage{amssymb}
\usepackage{amsfonts}
\usepackage{filecontents}
\usepackage{color}
\usepackage[normalem]{ulem} 
\usepackage{soul}
\usepackage{epstopdf}
\usepackage{nameref}
\usepackage{hyperref}

\hypersetup{hidelinks}
\usepackage{fancyhdr}
\usepackage{float}
\usepackage{gensymb}
\usepackage{physics}
\usepackage{upgreek}
\newcounter{fig}
\usepackage{lipsum}
\usepackage{natbib}
\usepackage{bibunits} 
\usepackage[usenames,dvipsnames,svgnames,table]{xcolor}
\usepackage{enumitem}
\usepackage{multirow}
\usepackage{upgreek}
\DeclareRobustCommand{\SkipTocEntry}[5]{}
\usepackage{titlesec}
\usepackage{indentfirst}
\defaultbibliography{Auger_paper_reference_1.bib}
\renewcommand{\bibliography}[1]{} 
\begin{document}
\begin{bibunit}[naturemag]

\renewcommand\thesection{\Roman{section}.}

\titleformat{\section}
  {\normalfont\bfseries\centering}
  {\thesection}{0.7em}{\MakeUppercase}

\titlespacing*{\section}{0pt}{4.2ex plus 1.2ex minus 0.6ex}{2.4ex plus 0.6ex minus 0.3ex}

\title{Structured Single-photon Metasource
}

\author{Jun-Yong Yan}
\thanks{These authors contributed equally to this work.}
\affiliation{Department of Electrical and Computer Engineering, National University of Singapore, Singapore, Singapore}

\author{Fang-Yuan Li}
\thanks{These authors contributed equally to this work.}
\affiliation{State Key Laboratory of Extreme Photonics and Instrumentation, Zhejiang University, Hangzhou, China}
\affiliation{College of Information Science and Electronic Engineering, Zhejiang University, Hangzhou, China}

\author{Zhou Zhou}
\thanks{These authors contributed equally to this work.}
\affiliation{Department of Electrical and Computer Engineering, National University of Singapore, Singapore, Singapore}

\author{Yue-Yao Mu}
\thanks{These authors contributed equally to this work.}
\affiliation{State Key Laboratory of Extreme Photonics and Instrumentation, Zhejiang University, Hangzhou, China}
\affiliation{College of Information Science and Electronic Engineering, Zhejiang University, Hangzhou, China}

\author{Hang-Yu Ge}

\affiliation{State Key Laboratory of Extreme Photonics and Instrumentation, Zhejiang University, Hangzhou, China}
\affiliation{College of Information Science and Electronic Engineering, Zhejiang University, Hangzhou, China}

\author{Severin Krüger}
\affiliation{Experimental Physics VI, Ruhr University Bochum, Bochum, Germany}

\author{Jianfeng Chen}
\author{Zhe Wang}
\author{Fulong Shi}
\author{Mengqi Liu}
\author{Haoye Qin}
\author{Ying Che}
\affiliation{Department of Electrical and Computer Engineering, National University of Singapore, Singapore, Singapore}

\author{Yu-Tong Wang}

\affiliation{State Key Laboratory of Extreme Photonics and Instrumentation, Zhejiang University, Hangzhou, China}
\affiliation{College of Information Science and Electronic Engineering, Zhejiang University, Hangzhou, China}

\author{Yunyan Zhang}
\email[Email to: ]{yunyanzhang@zju.edu.cn}
\affiliation{College of Integrated Circuits, Zhejiang University, Hangzhou, Zhejiang, China}

\author{Song Han}
\affiliation{College of Information Science and Electronic Engineering, Zhejiang University, Hangzhou, China}

\author{Zongyin Yang}
\affiliation{College of Information Science and Electronic Engineering, Zhejiang University, Hangzhou, China}

\author{Chaoyuan Jin}
\affiliation{ZJU-Hangzhou Global Scientific and Technological Innovation Center, Zhejiang University, Hangzhou, China}

\author{Huiyun Liu}
\affiliation{Department of Electronic and Electrical Engineering, University College London, London WC1E 7JE, UK}

\author{Arne Ludwig}
\affiliation{Experimental Physics VI, Ruhr University Bochum, Bochum, Germany}

\author{Feng Liu}
\email[Email to: ]{feng\_liu@zju.edu.cn}
\affiliation{State Key Laboratory of Extreme Photonics and Instrumentation, Zhejiang University, Hangzhou, China}
\affiliation{College of Information Science and Electronic Engineering, Zhejiang University, Hangzhou, China}

\author{Cheng-Wei Qiu}
\email[Email to: ]{chengwei.qiu@nus.edu.sg}
\affiliation{Department of Electrical and Computer Engineering, National University of Singapore, Singapore, Singapore}
\affiliation{National University of Singapore Suzhou Research Institute, Suzhou, China}
\affiliation{Department of Physics, National University of Singapore, Singapore, Singapore}

\begin{abstract}
\textbf{
Structured quantum light is crucial for high-dimensional quantum information processing, yet its direct generation from quantum emitters remains challenging due to their intrinsic locality and omnidirectional radiation. Metasurfaces have been adopted for quantum-light wavefront shaping, typically in cascaded or stacked configurations that suffer from low efficiency and limited resolution. Here, we demonstrate a semiconductor metasource that directly embodies single quantum dots in a nonlocal GaAs metasurface. Spontaneous emission from quantum dot is efficiently funneled into an extended quasi-bound-state-in-the-continuum mode while sustaining strong mode-emitter overlap. A lateral core-barrier heterostructure tunes mode volume and spatial distribution to balance Purcell enhancement and holographic resolution. Using spatially modulated geometric phase, our compact metasource enables deterministic generation of diverse single-photon radiation patterns, including orbital-angular-momentum beams and holographic images. Our work brings versatile single-photon wavefront control into the nanoscale cavity quantum electrodynamics regime, offering a scalable route toward integrated sources of structured quantum light.
}

\end{abstract}

\maketitle

\addtocontents{toc}{\protect\SkipTocEntry}
\section{Introduction}

Solid-state quantum emitters play a pivotal role in scalable photonic quantum technologies. After decades of rapid progress, these systems can now generate high-performance non-classical light on demand~\cite{Aharonovich2016,Daggett2026,Tomm2021a,Jeannic2022,Ding2023,Liu2024}, with applications in quantum computing~\cite{Maring2024}, quantum communication~\cite{Zou2025} and metrology~\cite{Muller2017}. In parallel, flat-optics wavefront-engineering platforms, such as metasurfaces~\cite{Dorrah2022}, have enabled powerful control over optical fields, including beam steering~\cite{Iyer2022}, holography~\cite{Zheng2015,Xiong2023}, orbital-angular-momentum generation~\cite{Arbabi2015} and arbitrary wavefront shaping~\cite{Zeng2025}. Their planar geometry, compact footprint and compatibility with standard nanofabrication technologies make them an attractive route towards integrated photonic functionalities. Extending such capabilities to the quantum regime to realize spatially structured quantum light sources is highly desirable, as structured quantum light can enlarge the accessible Hilbert space for high-capacity information encoding and offers new opportunities for high-resolution quantum imaging~\cite{Solntsev2021,Forbes2021,Forbes2025}.

\begin{figure*}
\refstepcounter{fig}
    \includegraphics[width=0.9\textwidth]{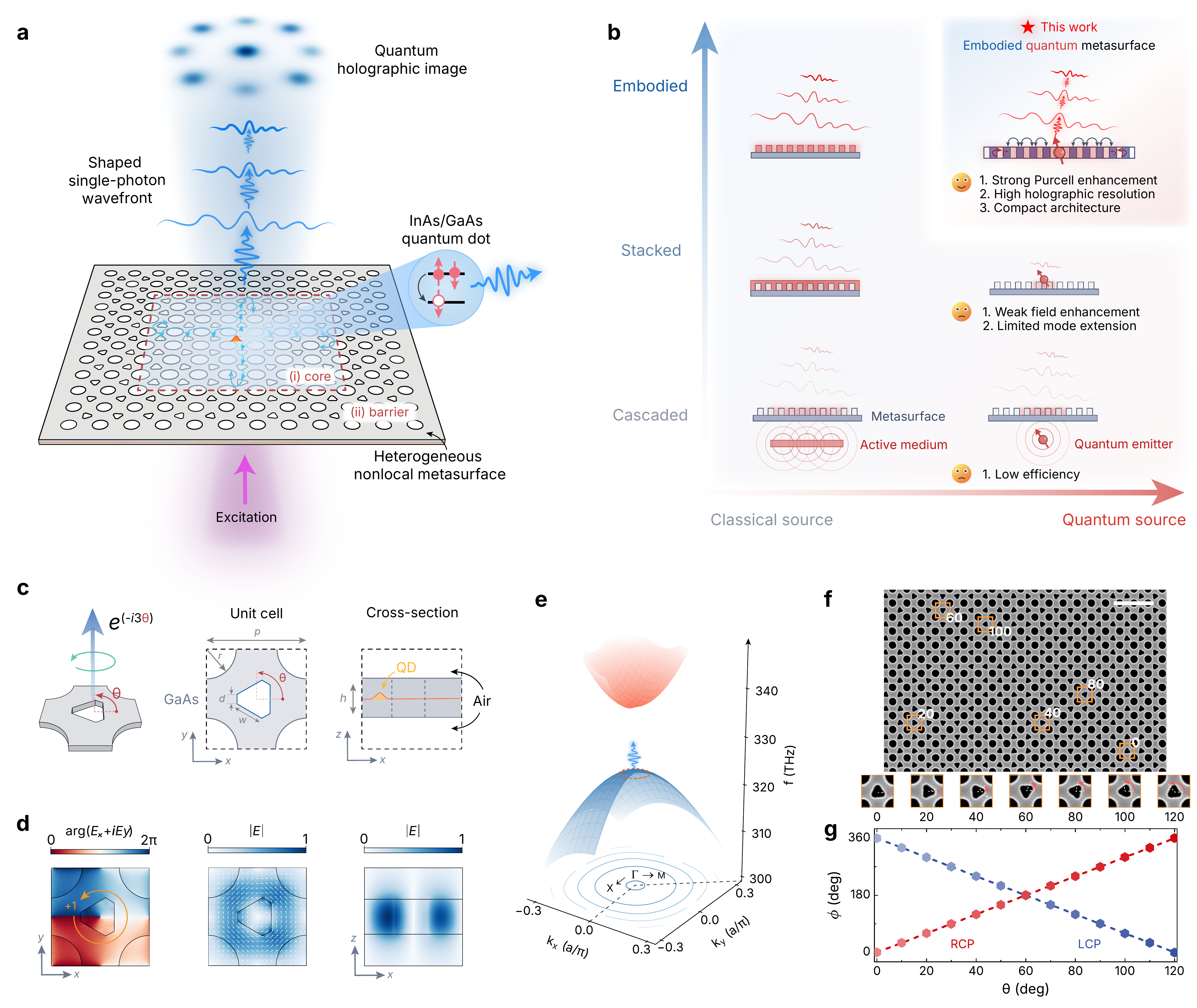} 
    \caption{\textbf{Arbitrarily structured single photons from a QD-coupled nonlocal heterogeneous metasurface.}
    \textbf{a}, Schematic of a single quantum emitter coupled to a nonlocal heterogeneous metasurface. A single QD at the cavity center is optically excited, and the emitted photons are funneled into the mode and shaped into a designed wavefront.
    \textbf{b}, Comparison of our embodied quantum metasurface with other active metasurface platforms. Unlike existing structured quantum-light sources based on cascaded or stacked metasurface-emitter integration, our approach waives such integration while achieving strong Purcell enhancement and high holographic resolution in a single-layer device.
    \textbf{c}, Schematic of a unit cell with perturbation rotation angle $\uptheta$, inducing a phase shift $\phi = -3\uptheta$ in the emitted LCP component. The geometric parameters are $p=382$~nm, $r=118$~nm, $h=140$~nm, $w=115$~nm and $d=39$~nm.
    \textbf{d}, Simulated phase distribution of the LCP component (left) and electric-field distribution (middle and right) of the unit-cell eigenmode.
    \textbf{e}, Simulated photonic band structure, with the quasi-BIC at the Brillouin-zone center (blue) selected as the operating mode.
    \textbf{f}, Top-view SEM image of the fabricated nonlocal metasurface. Lower inset: seven representative unit cells with different rotation angles. Scale bar, 1~$\upmu$m.
    \textbf{g}, Induced phase retardance $\phi$ of the LCP (blue) and RCP (red) components as a function of perturbation rotation angle $\uptheta$.}
\label{Fig:principle}
\end{figure*}

However, a quantum emitter is intrinsically spatially localized, and its spontaneous emission is usually omnidirectional. As a result, existing structured quantum-light sources usually rely on cascaded~\cite{Li2020,Huang2019,Bao2020,Li2023} or stacked~\cite{Rong2020,Komisar2023} metasurface configurations (Fig.~\ref{Fig:principle}b). Such approaches provide weak emitter-mode overlap, which limits source efficiency and lacks light-matter interaction enhancement. Furthermore, a quantum emitter stacked on a metasurface can typically excite only a few nearby meta-atoms, resulting in poor mode extension and thus limited wavefront-shaping resolution.

Recently, nonlocal metasurfaces~\cite{Zhang2022a,Shastri2023,Qin2025}, in which excitation at a given point affects the response over an extended spatial region, have emerged as a promising platform for engineering customized lasing properties~\cite{Duan2024,Zeng2025}. Meanwhile, embodied metasurface architectures, with the active medium forming an integral part of the metasurface rather than being externally added, can further improve emitter-mode overlap and have been demonstrated to reduce lasing thresholds~\cite{Chen2023}. However, the weak lateral confinement of the delocalized mode conflicts with the small mode volume required for Purcell enhancement. Despite the success of nonlocal classical metasources, a genuine metasource of structured quantum light that simultaneously achieves high holographic resolution and strong Purcell enhancement has remained elusive.

Here, we propose and experimentally demonstrate a compact planar semiconductor metasurface that co-designs enhanced light-matter interaction and wavefront formation within a single 140-nm-thick layer. By directly embedding quantum dots in a nonlocal heterogeneous metasurface, the overlap between the emitter and the resonant mode is maximized, while spontaneous emission is efficiently funneled into the extended nonlocal mode responsible for wavefront shaping. A lateral heterostructure design consisting of core and photonic-barrier regions with different air-hole sizes creates a photonic 'wall' and enables flexible control over both mode volume and wavefront-shaping resolution. Fabricated in a single etching step, our metasource device exhibits substantial Purcell enhancement ($F_p=10.1$) and directly generates coherent superpositions of distinct spatially structured modes in a high-purity single-photon state ($g^2(0)=0.018$). The emitted photons form customized orbital-angular-momentum (OAM) beams, as well as single-spot and multi-spot holographic patterns, without any external meta-optical elements. Our work brings planar wavefront engineering into the regime of nanoscale cavity quantum electrodynamics (cQED), providing a compact, scalable and fabrication-friendly route towards metasources of structured quantum light for next-generation high-dimensional photonic quantum technologies.

\begin{figure*}
\refstepcounter{fig}
    \includegraphics[width=0.76\textwidth]{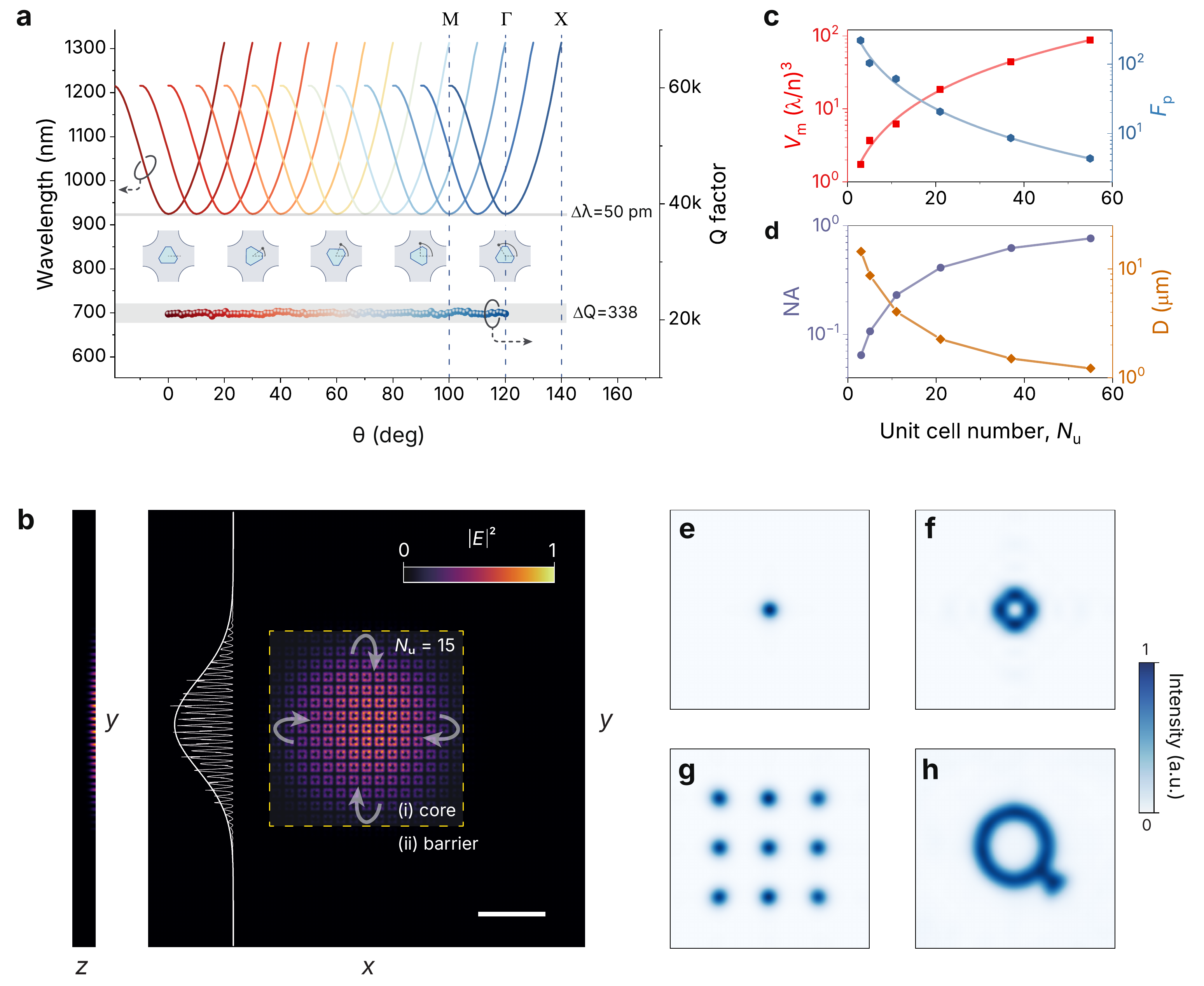} 
    \caption{\textbf{Design and optimization of the nonlocal heterogeneous metasurface.}
    \textbf{a}, Eigenmode band structure (top) and quality factor (bottom) versus perturbation rotation angle $\uptheta$ from 0 to 120~deg.
    \textbf{b}, Simulated electric-field intensity distribution of the heterostructure, showing effective longitudinal and lateral confinement in the slab core region. By varying the unit-cell number in the core region ($N_u$), the mode volume and cavity NA are flexibly tuned. The intensity profile along $y$ is extracted and fitted with a Gaussian function (white line). Scale bar, 2~$\upmu$m.
    \textbf{c,d}, Calculated cavity mode volume $V_m$, maximum Purcell factor $F_p$, effective NA, and corresponding holographic resolution $D$ as functions of the core-region unit-cell number $N_u$. Increasing $N_u$ increases $V_m$ and NA, improving holographic resolution at the cost of reduced $F_p$.
    \textbf{e-h}, Numerically predicted single-photon holographic images of one spot (\textbf{e}), an OAM vortex spot with $l=1$ (\textbf{f}), nine spots (\textbf{g}), and the letter ‘Q’ (\textbf{h}).}
\label{Fig:simulation}
\end{figure*}

\addtocontents{toc}{\protect\SkipTocEntry}
\section{Working principle}

The working principle of the device is schematically illustrated in Fig.~\ref{Fig:principle}a. The nonlocal metasurface is formed by nanohole arrays patterned in a suspended GaAs membrane using electron-beam lithography and reactive-ion etching, followed by selective undercutting of the sacrificial $\rm{Al_{0.75}Ga_{0.25}}$As layer~\cite{Tiranov2023,Wang2025,Liu2018c} (see supplementary materials for fabrication details). Each unit cell comprises four quarter-circle air holes and a corner-truncated equilateral triangular hole (Fig.~\ref{Fig:principle}c). A layer of self-assembled InAs/GaAs quantum dots (QDs) is embedded at the center of the membrane~\cite{Warburton2013b}. A selected QD positioned at the metasurface center couples to the engineered nonlocal mode and serves as the single-photon emitter. When its emission is tuned into resonance with the high-quality-factor cavity mode, the Purcell effect enhances spontaneous emission and efficiently funnels the emitted single photons into the resonant mode. Through the engineered geometric-phase distribution, the metasurface then shapes this emission into a designed free-space wavefront, generating structured single-photon patterns. To simultaneously support a delocalized mode and maintain strong Purcell enhancement, we employ a heterogeneous design between the core and surrounding barrier regions, in which the sizes of the circular and triangular holes in the barrier region are scaled to 0.95 of those in the core (see Supplementary Note~\ref{Note: heterostructure}).

Figure~\ref{Fig:principle}d presents the simulated electric-field intensity distribution (middle and right panel) and the phase profile of the LCP electric-field component (left panel). The field exhibits a doughnut-shaped profile confined within the slab, facilitating efficient spatial overlap with the embedded QDs. The left-circularly polarized phase presents a $2\pi$ azimuthal winding, carrying a topological charge of 1 and providing robustness against moderate geometric perturbations~\cite{Zhen2014}. The calculated band structure is shown in Fig.~\ref{Fig:principle}e, where the quasi-bound state in the continuum (quasi-BIC, blue) is selected as the operating mode. In the absence of symmetry-breaking perturbations (i.e., without triangular holes), the structure supports a nonlocal BIC characterized by strongly suppressed out-of-plane radiation~\cite{Lee2012,Wang2020a}. Introducing truncated triangular holes imposes a controllable geometric-phase perturbation that breaks the in-plane inversion $C_{4}$ symmetry. Thus, the right- and left-circularly polarized (RCP and LCP) components of the emitted light acquire opposite helicity-dependent phase shifts of $+3 \uptheta$ and $-3 \uptheta$, respectively~\cite{Overvig2020,Xie2021}, where $\uptheta$ denotes the triangle orientation angle (see Fig.~\ref{Fig:principle}c and Supplementary Note~\ref{Note: Geometric phase}). The required spatial phase modulation map $\phi(x,y)$ is calculated using the angular spectrum method combined with gradient-descent optimization~\cite{So2023} and is implemented by spatially varying triangle orientation angle $\uptheta(x,y)$. A representative device is shown in the top-view scanning electron microscopy (SEM) image in Fig.~\ref{Fig:principle}f. Fig.~\ref{Fig:principle}g presents the simulated acquired geometric-phase retardance $\phi$ of emitted LCP and RCP light for different perturbation angle $\uptheta$, showing a perfect linear relationship.

\addtocontents{toc}{\protect\SkipTocEntry}
\section{Design and optimization of quantum metasource}

Preserving the intrinsic resonance characteristics of the quasi-BIC mode under spatial phase modulation is essential for realizing stable single-photon wavefront shaping without compromising microcavity performance. In particular, the resonance wavelength, quality (Q) factor, and field distribution should remain nearly invariant when introducing pixel-wise rotational perturbations. As shown in Fig.~\ref{Fig:simulation}a, the simulated photonic band structure along M$-\Gamma-$X points and Q factor exhibit only minimal variations, with a band-edge wavelength deviation $\Delta\lambda$ of 50~pm (0.005\%) and a Q-factor deviation $\Delta Q$ of 338 (1.6\%). The average Q factor is 21097, which could be flexibly tuned by perturbation strength (Supplementary Note~\ref{Note: structural parameters}). These results confirm the superior topological resilience of the quasi-BIC mode to the engineered perturbations than other modes~\cite{Rong2020}.

For quantum photonic devices, enhancing light–matter interaction through cQED effects is essential for efficient photon extraction, improved robustness against charge noise and incoherent phonon sidebands~\cite{Liu2018c,Iles-Smith2017b,Grange2017a}. This enhancement of spontaneous emission is usually quantified by the Purcell factor~\cite{Purcell1946}:

\begin{equation}
F_{\mathrm p}=
\frac{3}{4\pi^2}\times\,
\frac{Q}{V_m}\times\,
\frac{1}{1+4\left(\Delta/\kappa\right)^2}\,\times\,
\frac{\left|\boldsymbol{\mu}\cdot \boldsymbol{E}(\mathbf r)\right|^2}{|\boldsymbol{\mu}|^2\left|\boldsymbol{E}_{\max}\right|^2},
\label{eq:Purcell}
\end{equation}

where $Q$ and $V_m$ represent the cavity quality factor and the mode volume in units of $(\lambda/n)^3$. $\Delta$ denotes the emitter–cavity detuning and $\kappa$ the full width at half maximum (FWHM) of the cavity mode, both expressed in frequency units. The vectors $\boldsymbol{\mu}$, $\boldsymbol{E}(\mathbf{r})$, and $\boldsymbol{E}_{\max}$ represent the emitter transition dipole moment, the local electric field at the emitter position, and the cavity-mode electric field at the intensity maximum, respectively. The Purcell enhancement therefore depends not only on a high Q factor but also on a small mode volume. Figure~\ref{Fig:simulation}b shows the field-intensity distribution of the heterostructure metasurface for $N_u=15$, where the mode is effectively confined in both the longitudinal and lateral directions. We further simulate the cavity mode volume and the corresponding maximum Purcell factor as functions of $N_u$ (Fig.~\ref{Fig:simulation}c). The effective numerical aperture of metasurface, $\mathrm{NA}=\sin\Bigl(\arctan\bigl(\frac{N_u\cdot p}{2f}\bigr)\Bigr)$, and the corresponding holographic resolution, $\mathrm{D}=\frac{\lambda}{2\mathrm{NA}}$, are also evaluated (Fig.~\ref{Fig:simulation}d), with the device focal length $f$ fixed at 10~$\upmu$m.

We note that although a trade-off exists between the achievable Purcell factor and holographic resolution, our architecture allows flexible tuning of the effective mode volume and NA, enabling Purcell enhancement and wavefront-shaping functionality to be balanced for specific design requirements. Figures~\ref{Fig:simulation}e-h present simulated ideal holographic images of a single spot, a vortex spot carrying OAM $l=1$, nine spots, and the letter 'Q', corresponding to $N_u=5$, 11, 71, and 91, respectively, with maximum achievable Purcell factors of 103, 61, 3, and 2.

\begin{figure*}[!htbp]
\refstepcounter{fig}
    \includegraphics[width=0.9\textwidth]{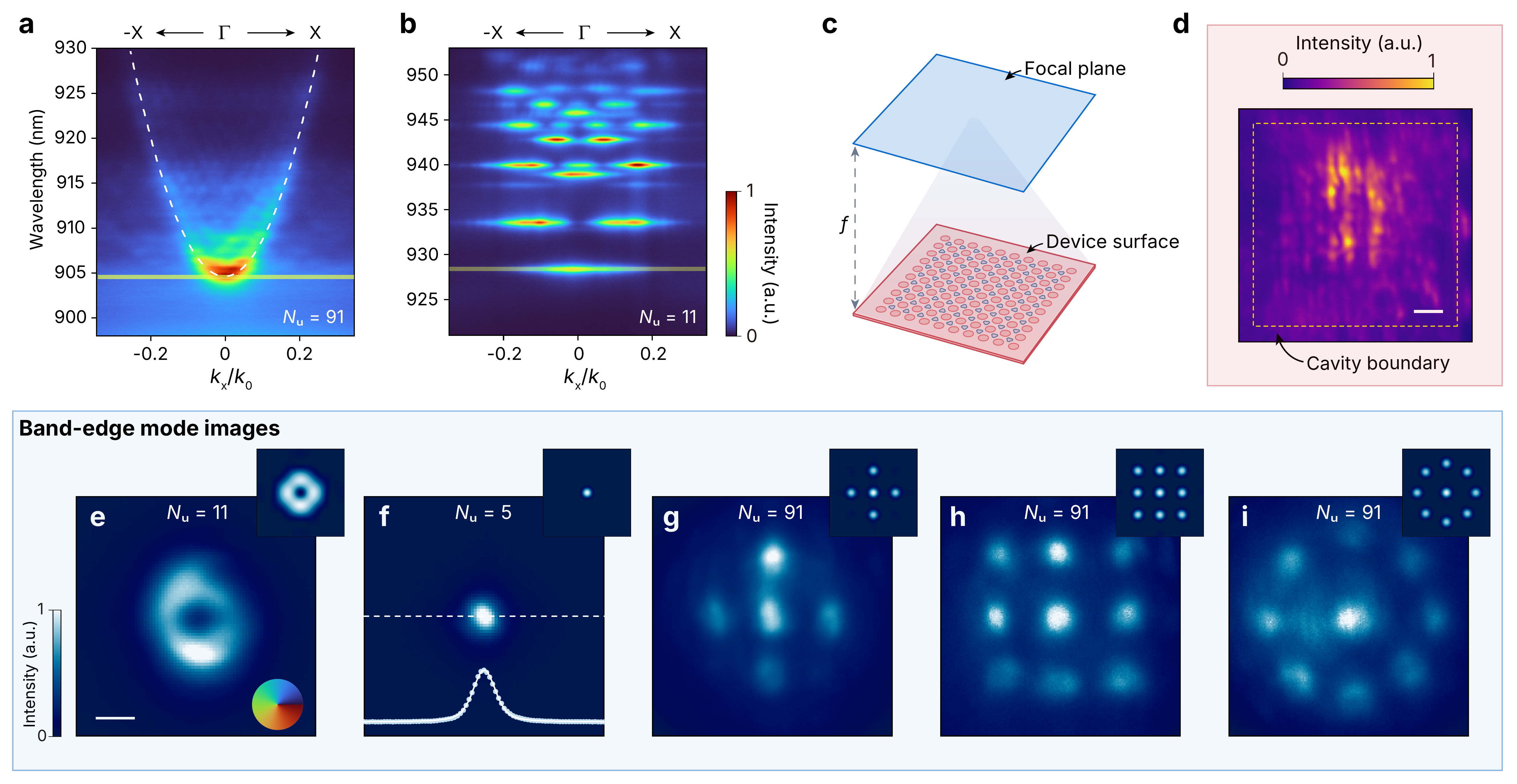} 

    \caption{\textbf{Observation of structured light from the nonlocal heterogeneous metasurface.}
    \textbf{a,b}, Photonic band structures of the metasurface with $N_u=91$ (\textbf{a}) and $N_u=11$ (\textbf{b}), measured by momentum-resolved spectroscopy. A 0.4~nm bandpass filter (yellow shaded area) selects the band-edge component for holographic imaging. White dashed lines: simulated band structures.
    \textbf{c}, Schematic of the imaging planes. Blue: designed focal plane. Red: device surface plane.
    \textbf{d}, Measured intensity distribution at the device surface plane.
    \textbf{e--i}, Measured intensity distributions at the designed focal plane. Five devices with different spatial phase profiles generate an OAM vortex with $l=1$ (\textbf{e}), a single spot (\textbf{f}), five spots (\textbf{g}), a nine-spot square (\textbf{h}), and a circular array (\textbf{i}). Upper-right insets: simulated intensity distributions. Lower-right inset of \textbf{e}: phase map of the vortex spot. Lower inset of \textbf{f}: intensity profile along $y=0$. Scale bar, 5~$\upmu$m.}
    
\label{Fig:cavity}
\end{figure*}

\addtocontents{toc}{\protect\SkipTocEntry}
\section{Structured classical light from nonlocal heterogeneous metasurface}

To validate the proposed quantum light metasource, we fabricated the device in a molecular-beam-epitaxy-grown GaAs membrane containing InAs/GaAs QDs~\cite{Michler2000a}. The thickness ($h$), lattice constant ($p$), and hole radius ($r$) of the metasurface are 140, 382, and 118~nm, respectively, to match the QD emission near 925~nm (see Supplementary Note~\ref{Note: structural parameters}). The experimental setup for optical characterization is detailed in supplementary materials and schematically shown in Fig.~\ref{Extended Fig:setup}. To determine the band-edge frequency of the resonant mode, we measured the photonic band structures of two heterogeneous metasurface cavities with different $N_u$ using momentum-space photoluminescence (PL) spectroscopy. The QD ensembles in the devices are excited by an above-bandgap wide-field laser, which efficiently populates the cavity modes~\cite{Winger2009} and reveals their dispersion (Fig.~\ref{Fig:cavity}a). As $N_u$ decreases from 91 to 11, the finite-size effect~\cite{Chen2022,Contractor2022} quantizes the continuous dispersion into a set of discrete modes (Fig.~\ref{Fig:cavity}b). Unlike a conventional BIC that strongly suppresses radiation~\cite{Wang2020a} (Supplementary Note~\ref{Note: unperturbed BIC}), the perturbation-engineered quasi-BIC exhibits pronounced emission at $\Gamma$ point, which is beneficial for bright light-emitting devices. A bandpass filter is applied to spectrally select the band-edge component (yellow shaded area). As indicated in Fig.~\ref{Fig:cavity}c, we record images of both device surface (Fig.~\ref{Fig:cavity}d) and designed focal planes (Fig.~\ref{Fig:cavity}e-i). Five devices with different sizes and spatial geometric-phase profiles are characterized. Depending on the assigned phase distributions, the far-field distribution of band-edge emission was structured into an OAM vortex, a single spot, five spots, a nine-spot square, and a circular nine-spot array. For comparison, simulations of designed patterns are shown in the upper-right insets. We note that our architecture can also realize high-contrast spin-momentum-locked chiral emission (Supplementary Note~\ref{Note: chiral emission}), providing a potential route toward quantum chiral photonics.

\begin{figure*}[!htbp]
\refstepcounter{fig}
    \includegraphics[width=0.9\textwidth]{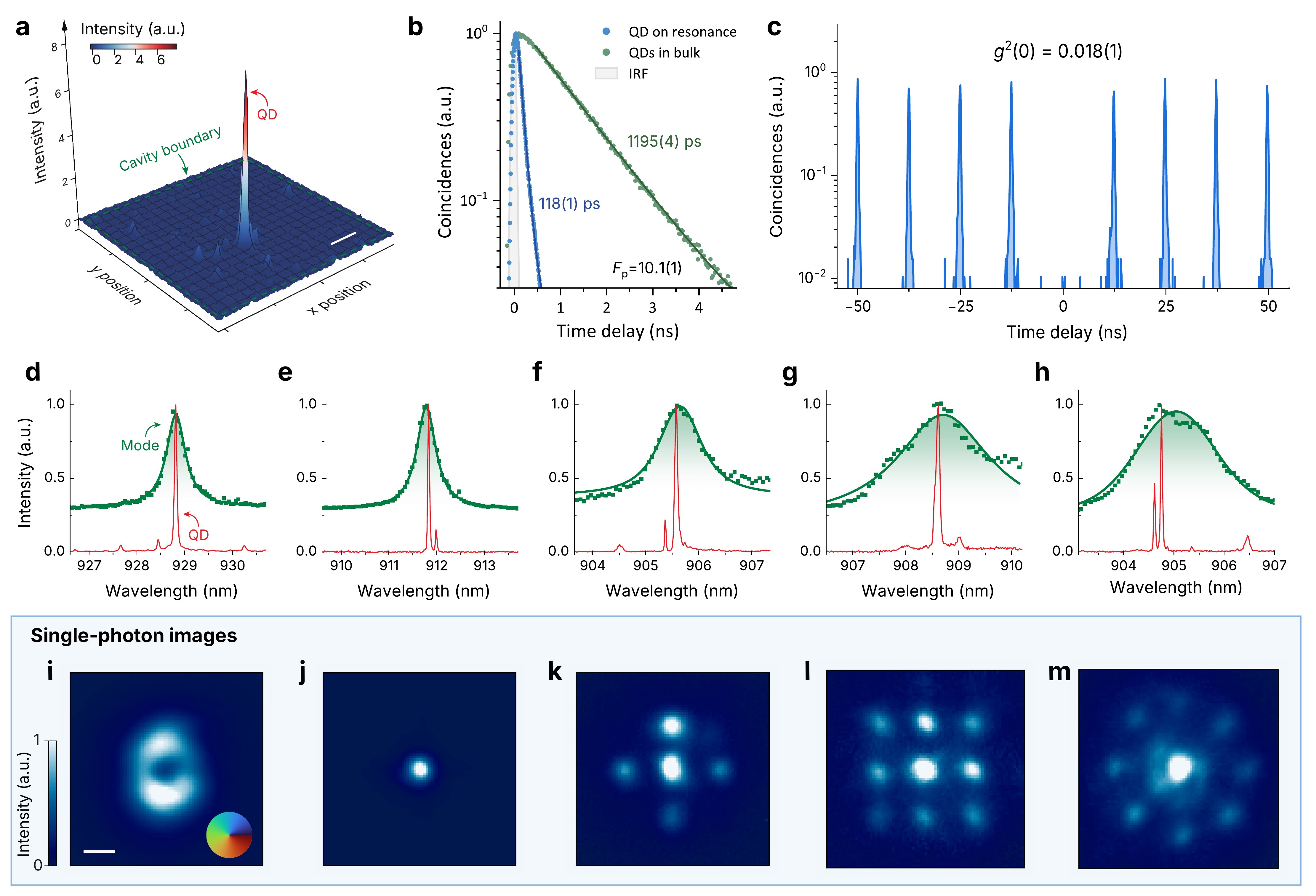} 
    \caption{\textbf{Tailoring and enhancing structured single-photon emission from quantum metasurfaces.}
    \textbf{a}, PL intensity map from confocal raster scanning under p-shell excitation, using a 0.4~nm detection window. Scale bar, 5~$\upmu$m.
    \textbf{b}, Radiative lifetimes of a QD coupled to the resonant mode (blue) and of QD ensemble in bulk (purple), showing a Purcell enhancement factor of 10.1(1). IRF, instrument response function.
    \textbf{c}, Log-scale second-order correlation function of the QD emission measured in a Hanbury--Brown and Twiss interferometer, yielding $g^{2}(0)=0.018(1)$.
    \textbf{d--h}, PL spectra of QDs (red) in the nonlocal metasurface and the corresponding resonant modes (green) from five devices with different holographic phase profiles. The cavity modes are fitted with a Voigt function.
    \textbf{i--m}, Corresponding single-photon holographic images.}
\label{Fig:QD}
\end{figure*}

\addtocontents{toc}{\protect\SkipTocEntry}
\section{Tailoring and enhancing structured single photons from quantum metasurfaces}

To operate the device as a quantum-light source, selective excitation of an individual QD is required. We therefore switch from wide-field to single-spot excitation and employ pulsed p-shell pumping. A two-dimensional raster confocal scan over the entire cavity region is performed (Supplementary Note~\ref{Note:  confocal raster scan}). The recorded PL intensity as a function of excitation position is shown in Fig.~\ref{Fig:QD}a, confirming that only a single QD is excited within each cavity. To efficiently couple the emitted single photons into the designed photonic mode, the emission wavelength of each QD is tuned into resonance with the band edge by precise temperature control. The radiative lifetime of a QD coupled to the cavity mode with $N_u=5$ is evaluated by time-resolved PL measurements. By comparison with the QD ensemble in bulk, a Purcell enhancement factor of 10.1(1) is extracted (Fig.~\ref{Fig:QD}b). The single-photon purity of the emitted light is verified by measuring the second-order autocorrelation function in a Hanbury-Brown and Twiss setup. The strongly suppressed coincidence at zero time delay ($1-g^{2}(0)=98.2\%$, Fig.~\ref{Fig:QD}c) indicates a negligible multiphoton probability and high-purity single-photon emission.

Figures~\ref{Fig:QD}d-h show the emission lines of the individual QDs (red) together with the corresponding cavity modes (green). As expected, the emitted single photons are shaped into distinct spatial distributions, as shown in Fig.~\ref{Fig:QD}i-m, in good agreement with both the numerical simulations and the experimentally measured mode characteristics in Fig.~\ref{Fig:cavity}. These clear quantum holographic images arise from efficient QD-mode coupling and confirm the successful population of the engineered nonlocal metasurface mode by the single-photon wavefunction.

\addtocontents{toc}{\protect\SkipTocEntry}
\section{Conclusions}

In summary, we propose and demonstrate arbitrarily structured single-photon emission from a compact QD-embedded nonlocal metasurface, featuring pronounced Purcell enhancement and high holographic resolution. The device consists of only a 140~nm-thick, single-step-etched membrane and is naturally compatible with scalable fabrication. The demonstrated tailorable OAM vortex beams and non-classical superposition states in the spatial-mode basis may provide valuable resources for emerging quantum technologies, including quantum random number generation~\cite{Herrero-Collantes2017}, high-resolution quantum imaging~\cite{Tenne2019}, high-dimensional quantum communication and computing~\cite{Cozzolino2019}, and multi-user quantum resource distribution. Furthermore, the embodied quantum metasurface concept is not limited to QDs, but could also be extended to other quantum-light platforms, such as transition metal dichalcogenides, colloidal and perovskite nanocrystals, solid-state defects, and nonlinear crystals.

Future work may explore a broader range of structured quantum-light states and effects, including quantum optical skyrmions~\cite{Ma2025,Cheng2026}, self-healing or perfect vortex beams~\cite{Shen2022,Ahmed2023}, generalized multi-port quantum interference~\cite{Yousef2025}, and the generation of photonic spin-orbital entanglement~\cite{Stav2018} and quantum graph states~\cite{Santiago-Cruz2022}. Our work paves the way toward flexibly customizable, miniaturized, and multifunctional quantum photonic devices for next-generation high-dimensional quantum technologies.

\newpage
\clearpage

\addtocontents{toc}{\protect\SkipTocEntry}
\section*{Acknowledgements}
We thank Prof. Limin Tong for the equipment support. We acknowledge the Micro/Nano Fabrication Center at Zhejiang University for their facility support and technical assistance. We also acknowledge the National Supercomputer Center in Guangzhou for computational support.

\textbf{Funding:} This work was supported by the National Key Research and Development Program of China (2023YFF0613600). Y.Z. and F.L. acknowledge support from the National Key Research and Development Program of China (No. 2023YFB2806000, 2022YFA1204700), the National Natural Science Foundation of China (No. U21A6006, 62075194, U24A20318, 62474163), Zhejiang Provincial Natural Science Foundation of China (No. LR25F040001, LZ24F040001), the Special Fund for Innovative Development of Hangzhou Chengxi Science and Technology Innovation Corridor (No. 226-2024-00050). C.-W.Q. acknowledged the financial support by the Ministry of Education, Republic of Singapore (Grant No.: A-8002152-00-00, A-8002458-00-00, A-8003643-00-00), and the Competitive Research Program Award (NRF-CRP26-2021-0004, NRF-CRP30-2023-0003) from the National Research Foundation, Prime Minister's Office, Singapore.


\addtocontents{toc}{\protect\SkipTocEntry}
\section*{Competing Interests}
The authors declare no competing interests.

\addtocontents{toc}{\protect\SkipTocEntry}
\section*{Data and materials availability}

All data are available in the main text or the supplementary materials.

\putbib 

\end{bibunit}

\begin{bibunit}[naturemag]

\setcounter{equation}{0}
\setcounter{figure}{0}
\setcounter{table}{0}
\setcounter{section}{0}

\renewcommand\thesection{\arabic{section}}
\renewcommand\thesubsection{\thesection.\arabic{subsection}}

\renewcommand{\figurename}{Figure}
\renewcommand{\tablename}{Table}
\renewcommand{\thefigure}{S\arabic{figure}}
\renewcommand{\thetable}{S\arabic{table}}
\renewcommand{\theequation}{S\arabic{equation}}
\renewcommand{\bibnumfmt}[1]{[S#1]}
\renewcommand{\citenumfont}[1]{S#1}

\renewcommand{\thesection}{\arabic{section}:}
\titleformat{\section}
  {\normalfont\bfseries\raggedright\LARGE}
  {\thesection}{1.0em}{}

\setcounter{subsection}{0}
\renewcommand{\thesubsection}{\arabic{subsection}}

\makeatletter
\renewcommand{\p@subsection}{}
\makeatother

\titleformat{\subsection}
  {\normalfont\bfseries\raggedright\Large}
  {Note~\thesubsection:}{0.6em}{}

\newcommand{\addtosi}[1]{%
  \addcontentsline{stoc}{section}{\protect\numberline{\thesection}#1}
}

\newcommand{\addsubtosi}[1]{%
  \addcontentsline{stoc}{subsection}{Note~\arabic{subsection}:~#1}
}

\newcommand{\addsubtosiunnumbered}[1]{%
  \addcontentsline{stoc}{subsection}{#1}
}

\titlespacing*{\subsection}{0pt}{1.2ex plus .2ex minus .2ex}{0.8ex}
\titlespacing*{\section}{0pt}{1.5ex plus .2ex minus .2ex}{1.0ex}

\setlength{\parindent}{2em}
\setlength{\parskip}{0pt}

\onecolumngrid
\newpage

\begin{Large}
\begin{center}
\textbf{Supplementary Materials for \\
Structured Single-photon Metasource}
\end{center}
\end{Large}

\setcounter{page}{1}

\begin{center}

Jun-Yong Yan $et~al$. \\
Corresponding authors: Yunyan Zhang, yunyanzhang@zju.edu.cn; 
Feng Liu, feng\_liu@zju.edu.cn;
Cheng-Wei Qiu, chengwei.qiu@nus.edu.sg

\end{center}

\clearpage
\newpage

\section*{Contents} 
\begingroup
\setcounter{tocdepth}{2}
\makeatletter
\@starttoc{stoc}
\makeatother
\endgroup

\section*{Materials and Methods } 
\addcontentsline{stoc}{section}{Materials and Methods}

\subsection*{Numerical simulations}
\addsubtosiunnumbered{Numerical simulations}
\label{Note: Numerical simulations}
All numerical simulations are performed using the finite element method (FEM). The GaAs slab is modeled as a lossless dielectric at the design wavelength (925~nm), with a refractive index of 3.5. For unit-cell simulations, Bloch boundary conditions are applied in the in-plane ($x$ and $y$) directions, while perfectly matched layers (PMLs) are implemented along the out-of-plane ($z$) direction to absorb radiative fields. For simulations of the full heterostructure metasurface cavity, a hemispherical computational domain with scattering boundary conditions is used to suppress spurious reflections and ensure reliable extraction of far-field radiation properties~\cite{Han2023}. The cavity mode volume is defined as~\cite{Painter1999}:
\begin{equation}
V_m=\frac{\int \varepsilon(\boldsymbol{r})|\mathbf{E}(\boldsymbol{r})|^2 \mathrm{d}^3 \boldsymbol{r}}{\max \left[\varepsilon(\boldsymbol{r})|\mathbf{E}(\boldsymbol{r})|^2\right]},
\label{eq:Vm}
\end{equation}
where $\varepsilon(\boldsymbol{r})$ is the dielectric constant and $|\mathbf{E}(\boldsymbol{r})|$ is the electric-field amplitude. The integration domain covers the entire slab. The maximum available Purcell factor is obtained by assuming zero QD-cavity detuning and optimal dipole position and orientation. Under these conditions, Eq.~\ref{eq:Purcell} reduces to
\begin{equation}
F_{\mathrm{P,~max}}=\frac{3}{4 \pi^2}\frac{Q}{V_m}.
\end{equation}

\newpage
\clearpage
\subsection*{Device fabrication}
\begin{figure*}[!htbp]
\refstepcounter{fig}
    \includegraphics[width=0.7\textwidth]{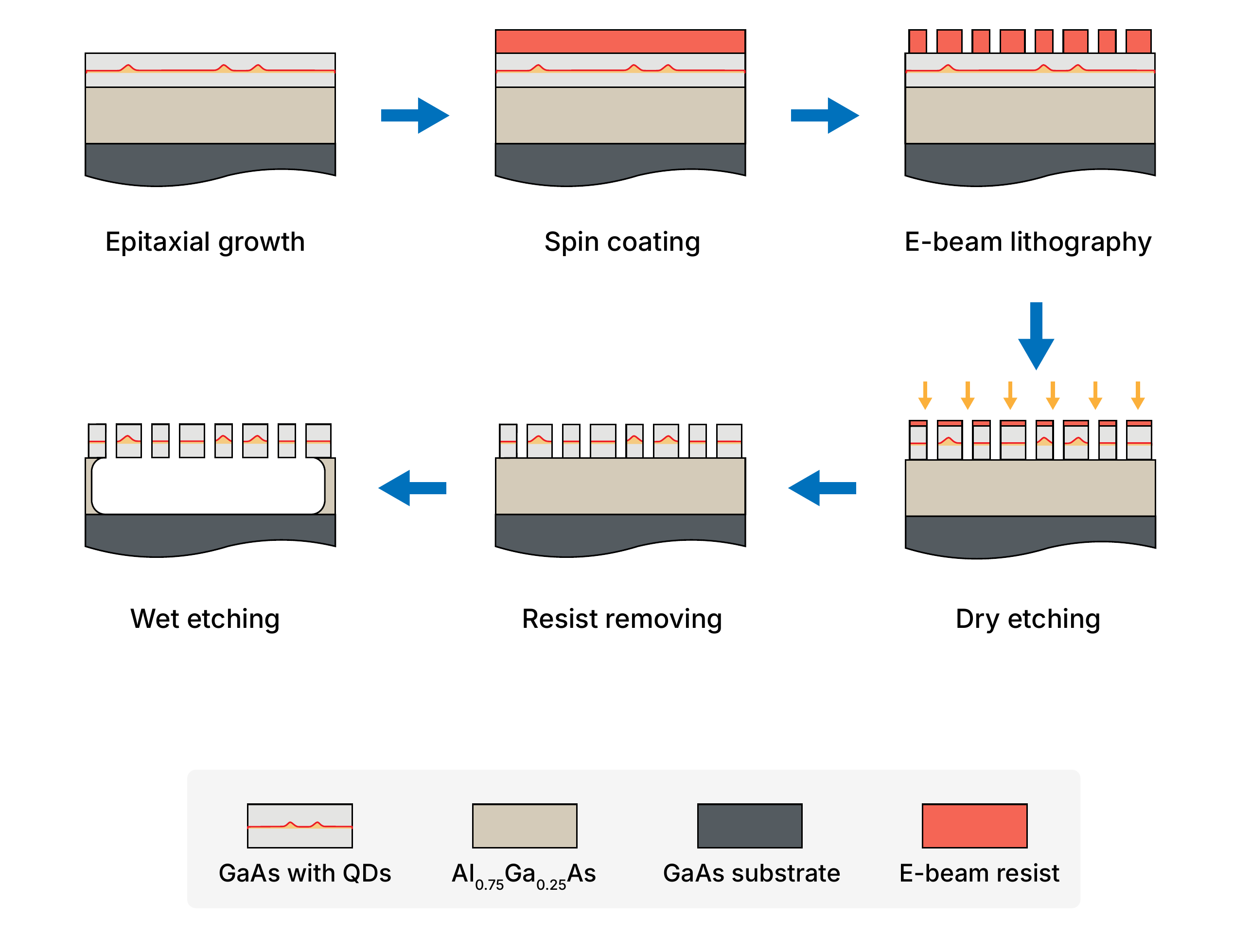} 
    \caption{\textbf{Main fabrication process of the sample.}
    The QD wafer is grown by molecular beam epitaxy. An electron-beam resist (AR-P 6200) is spin-coated onto the sample surface. The designed photonic structures are patterned by EBL and transferred into the membrane using ICP–RIE with a BCl$_3$/N$_2$ gas mixture. The residual resist is removed using N-methyl-2-pyrrolidone (NMP). To release the membrane, the sacrificial $\mathrm{Al_{0.75}Ga_{0.25}As}$ layer is selectively etched using a diluted hydrofluoric acid solution (HF:H$_2$O = 1:20).}
    
\label{Extended Fig:fab}
\end{figure*}

\addsubtosiunnumbered{Device fabrication}
\label{Note: Device fabrication}
The quantum metasurface device is fabricated following the process shown in Fig.~\ref{Extended Fig:fab}. The device wafer is grown by molecular beam epitaxy~\cite{Spitzer2023}. The devices are fabricated from a 140-nm-thick GaAs membrane containing a single layer of low-density self-assembled InAs/GaAs QDs at its center. The metasurface pattern is defined using a single CAD layout that includes both the square lattice of circular holes and the truncated equilateral triangular holes with predefined spatially modulated orientation angles. The pattern is transferred to an electron-beam resist by electron-beam lithography (EBL) in a single exposure step, followed by inductively coupled plasma reactive ion etching (ICP-RIE) to define the nanostructures in the GaAs membrane. Finally, the underlying $\rm{Al_{0.75}Ga_{0.25}}$As sacrificial layer is selectively removed by wet etching in a diluted hydrofluoric acid solution, resulting in a suspended membrane structure~\cite{Tiranov2023,Wang2025,Liu2018c}.

\newpage
\clearpage
\subsection*{Measurement techniques}
\begin{figure*}[!htbp]
\refstepcounter{fig}
    \includegraphics[width=1\textwidth]{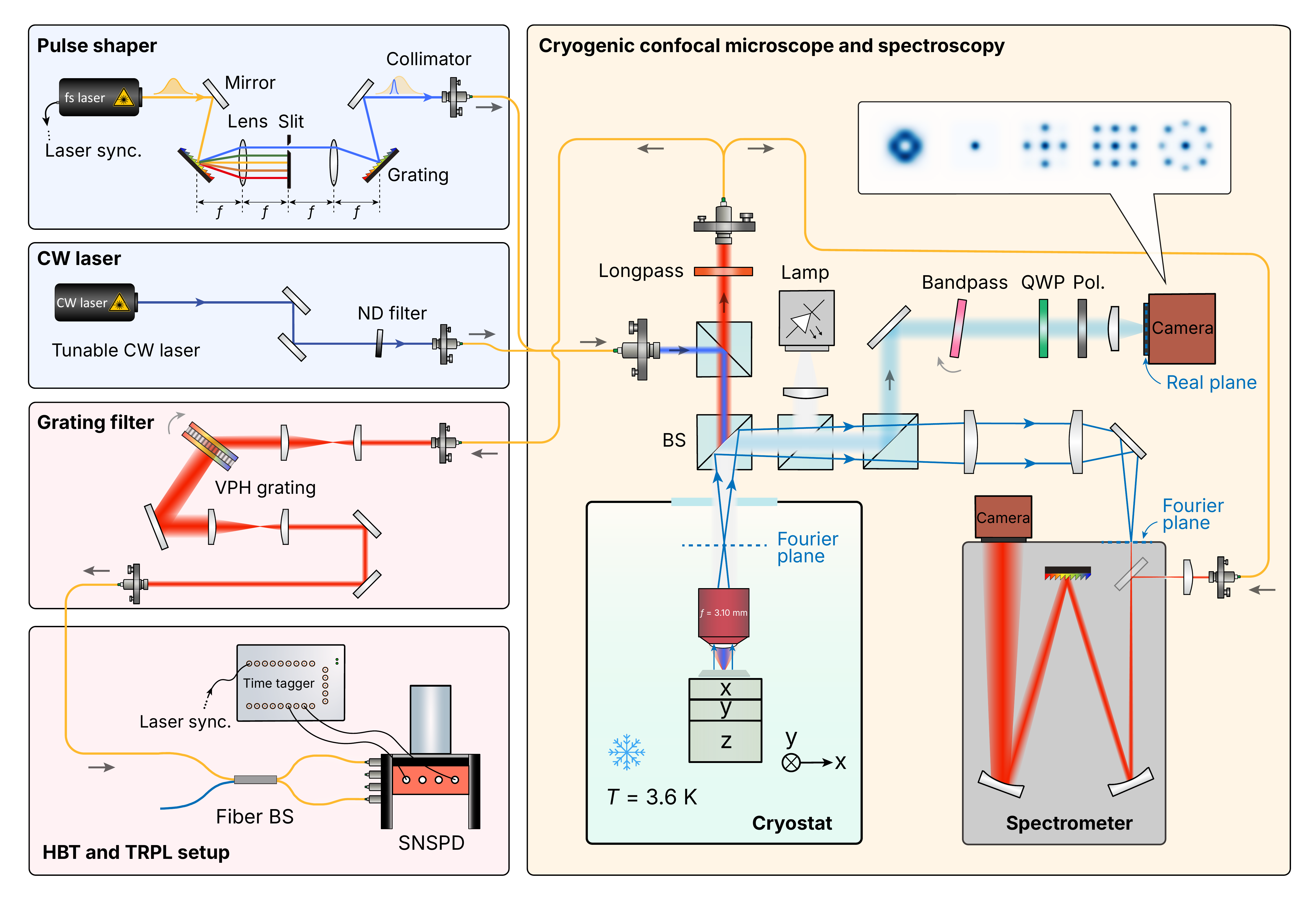} 
    \caption{\textbf{Schematic of the experimental setup for optical characterization.} The setup consists of five sections.
    \textbf{Pulse shaper}: A $4f$ femtosecond pulse shaper, composed of two lenses, two gratings, and a slit, generates tunable picosecond pulses with a spectral width of 0.8~nm.
    \textbf{CW laser}: A tunable continuous-wave laser is used for $p$-shell or above-band excitation.
    \textbf{Cryogenic confocal microscope and spectroscopy}: The device is mounted on an xyz nanopositioner inside the cryostat at 3.6~K. The excitation laser is focused onto the QD using an objective lens, and the emission light is collected for confocal or momentum-space spectroscopy. Holographic images are obtained after passing through a 0.4~nm bandpass filter, a quarter-wave plate (QWP), and a polarizer (Pol.), and are recorded by a sCMOS camera.
    \textbf{Grating filter}: A volume phase holographic (VPH) grating-based filtering module selects the QD emission line for temporal correlation measurements.
    \textbf{HBT and TRPL}: The filtered emission is directed to a multi-channel SNSPD system. Laser synchronization (sync.) and electrical signals are sent to a time tagger for time-stamping and correlation histogram acquisition.}
        
\label{Extended Fig:setup}
\end{figure*}

\addsubtosiunnumbered{Measurement techniques}
\label{Note: Measurement techniques}
A schematic of the experimental setup is shown in Fig.~\ref{Extended Fig:setup}~\cite{Yan2023}. The quantum metasurface device is mounted on an \emph{xyz} nanopositioner inside a closed-cycle cryostat operated at 3.6~K. A home-built $4f$ femtosecond pulse shaper is used to generate tunable picosecond pulses with a spectral width of 0.8~nm~\cite{Weiner2000}. The QD is excited either by the shaped picosecond pulses or by a tunable continuous-wave (CW) laser, depending on the measurement. The excitation laser is focused onto the QD through an objective lens ($f = 3.1$~mm, NA = 0.7). The emitted light is collected by the same objective lens and directed to a spectrometer through either free-space optics or an optical fiber for momentum-space or confocal spectral measurements, respectively. Holographic images are acquired after the emitted light passes through a 0.4~nm bandpass filter, a quarter-wave plate, and a polarizer, and are recorded by a scientific complementary metal–oxide–semiconductor (sCMOS) camera. For Hanbury-Brown and Twiss (HBT) and time-resolved photoluminescence measurements, a grating filter is used to select the QD emission line, which is then directed to the superconducting nanowire single-photon detectors (SNSPD). The laser synchronization signal and the electrical trigger signals from the SNSPD are sent to a time tagger for time stamping and correlation histogram generation.

\newpage
\clearpage
\section*{Supplementary Text}
\addcontentsline{stoc}{section}{Supplementary Text}

\subsection{Determination of the structural parameters of the metasurface}
\addsubtosi{Determination of the structural parameters of the metasurface}
\label{Note: structural parameters}

\begin{figure}[htbp]
\refstepcounter{fig}
    \includegraphics[width=1\textwidth]{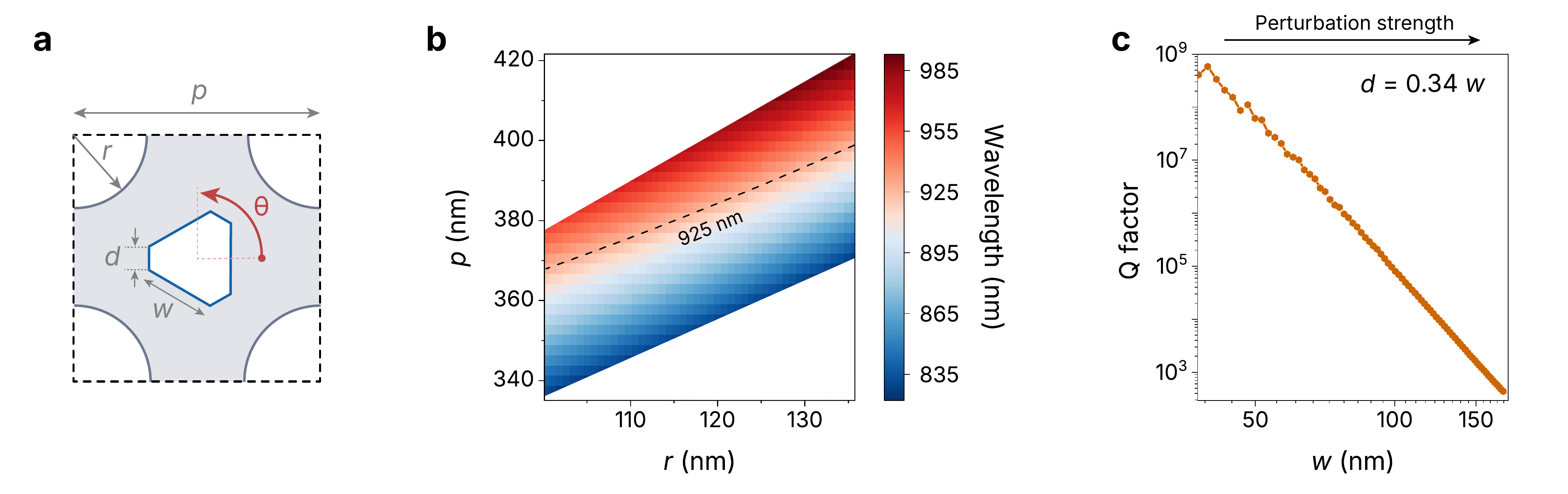} 
    \caption{\textbf{Determination of the structural parameters of the metasurface.}
    \textbf{a}, Schematic of the unit-cell geometry.
    \textbf{b}, Simulated resonant wavelength of the eigenmode. The dashed line indicates the target design wavelength, corresponding to the center wavelength of the QD ensemble emission.
    \textbf{c}, Quality factor of the eigenmode as a function of the perturbation strength. As the side length $w$ of the central triangle increases, the eigenmode evolves from a BIC to a quasi-BIC with a continuously tunable $Q$ factor.}
\label{Extended Fig:SFig paras scan}
\end{figure}

\begin{figure}[htbp]
\refstepcounter{fig}
    \includegraphics[width=0.8\textwidth]{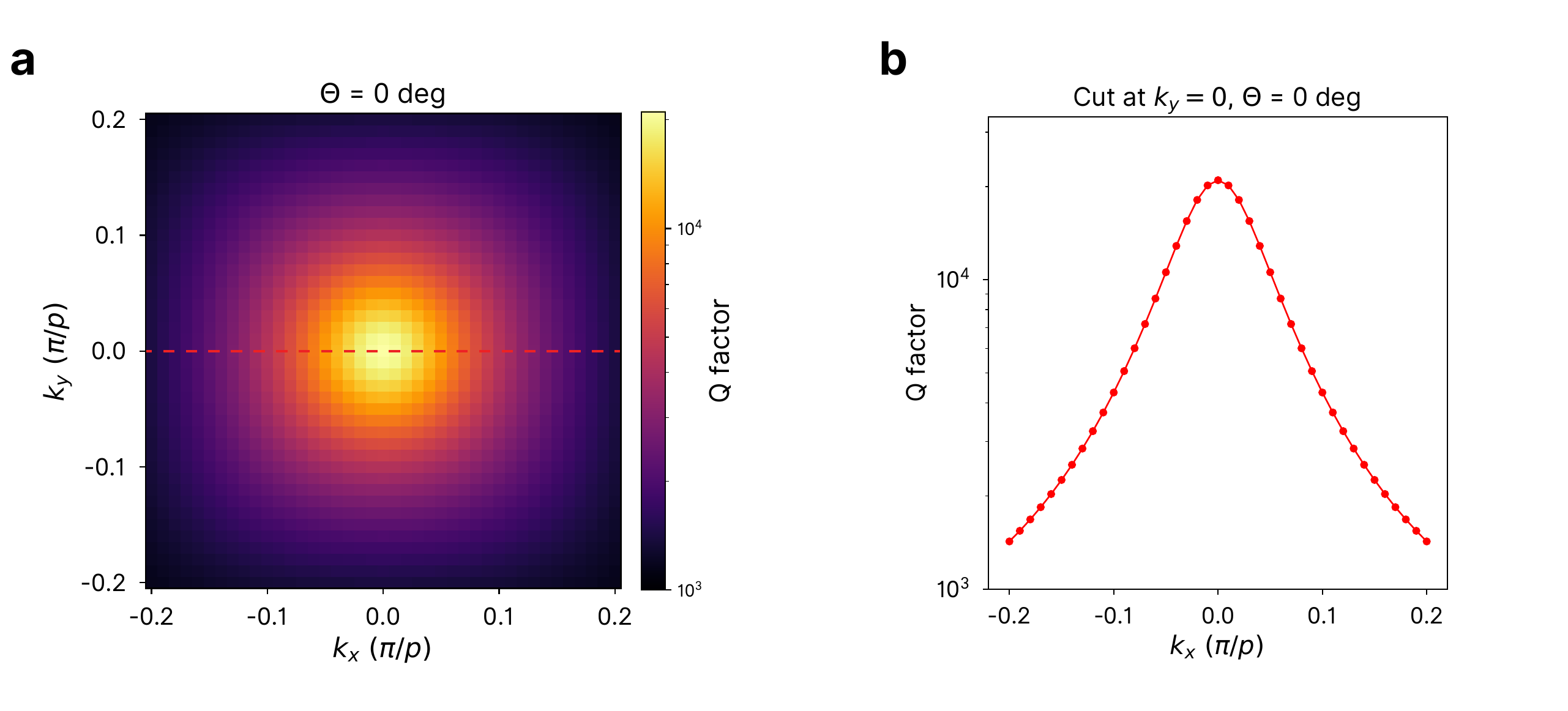} 
    \caption{\textbf{Momentum-space distribution of the quality factor.}
    \textbf{a}, Simulated quality factor in momentum space.
    \textbf{b}, Line cut at $k_y=0$ extracted from \textbf{a}.}
\label{Extended Fig:SFig Qfactor vs k}
\end{figure}
As shown in Fig.~\ref{Extended Fig:SFig paras scan}b, to match the resonant mode to the QD emission, we perform a two-dimensional scan of the metasurface unit-cell parameters ($p$ and $r$ in Fig.~\ref{Extended Fig:SFig paras scan}a). From these results, we determine the optimal geometry to be $p=382$~nm, $r=118$~nm, $h=140$~nm, $w=115$~nm, and $d=39$~nm. Figure~\ref{Extended Fig:SFig paras scan}c shows the quality factor ($Q$) as a function of $w$, with $d$ fixed at $0.34w$. As $w$ increases from 30 to 190, the $Q$ factor decreases from nearly infinity to 1000, indicating a transition from the BIC regime to the quasi-BIC regime. This result shows that the $Q$ factor can be continuously tuned by varying the perturbation strength $w$. We note that an extremely high $Q$ factor is not favorable for quantum light-source devices~\cite{Wang2026}, because it strongly suppresses out-of-plane radiation and thereby limits the source brightness. On the other hand, if the perturbation is too strong and the $Q$ factor becomes too low, the nonlocal character of the mode is compromised, leading to reduced mode extension and degraded holographic resolution. The chosen parameters yield a simulated $\Gamma$ point's $Q$ factor of 21097. Owing to fabrication imperfections and material losses, the experimentally measured $Q$ factor is about 1900. For the chosen parameters, we further simulate the momentum-space distribution of the $Q$ factor, as shown in Fig.~\ref{Extended Fig:SFig Qfactor vs k}a. Figure~\ref{Extended Fig:SFig Qfactor vs k}b shows a line cut at $k_y=0$. Moving away from the $\Gamma$ point, the quality factor decreases from approximately 20,000 to 1,000.

\subsection{Simulations of the core-barrier lateral heterostructure}
\addsubtosi{Simulations of the core-barrier lateral heterostructure}
\label{Note: heterostructure}

\begin{figure}[htbp]
\refstepcounter{fig}
    \includegraphics[width=0.95\textwidth]{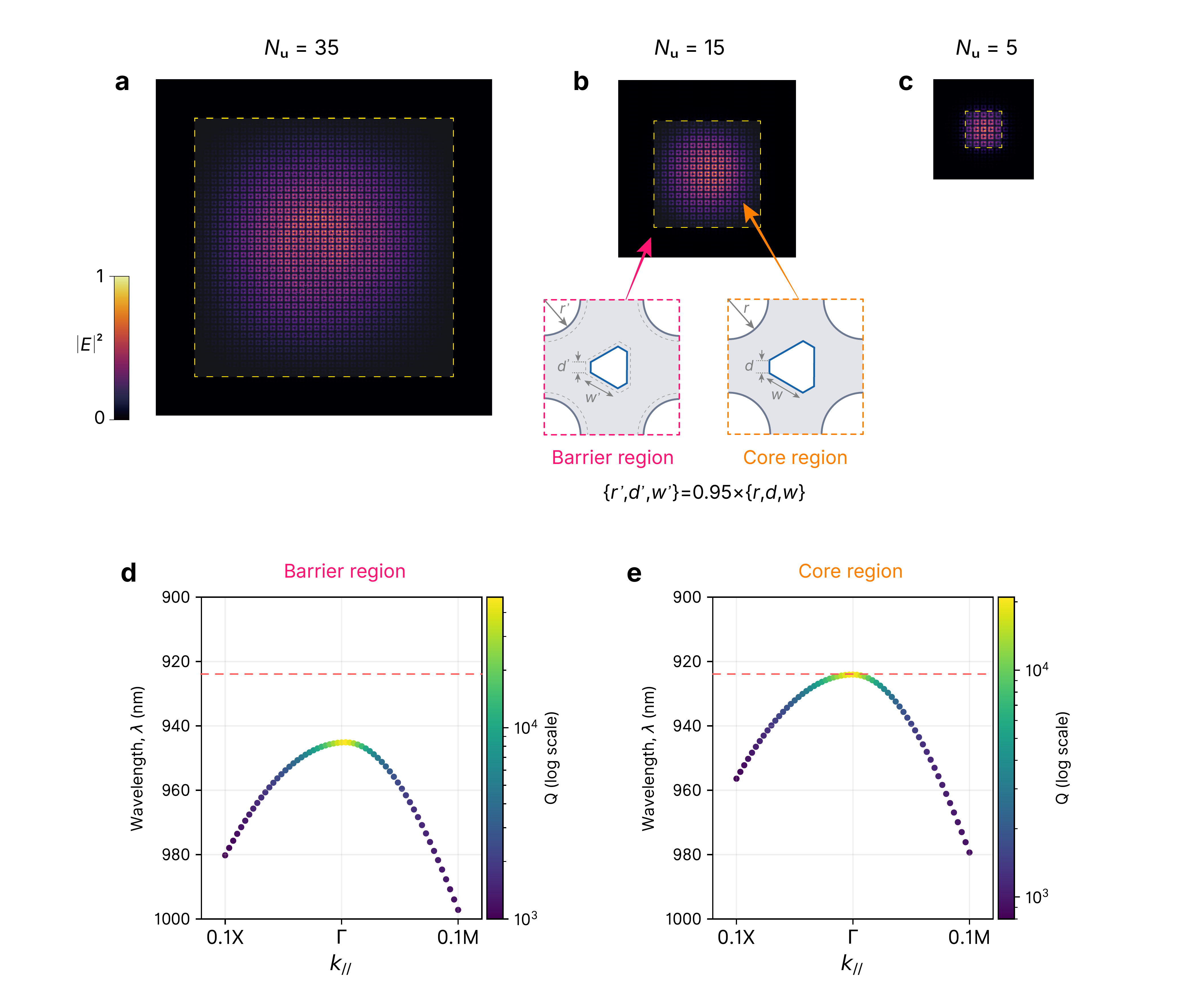} 
    \caption{\textbf{Simulated eigenmodes of the nonlocal heterogeneous metasurface.}
    \textbf{a-c}, Simulated resonant-mode profiles for core-region unit-cell numbers $N_u=35$ (\textbf{a}), 15 (\textbf{b}), and 5 (\textbf{c}). The heterogeneous cavity architecture enables a flexible trade-off between holographic resolution and Purcell enhancement. Inset: schematic illustration of the unit-cell geometries in the barrier and core regions. \textbf{d} and \textbf{e}, Band structures of the barrier (\textbf{d}) and core (\textbf{e}) regions, with the Q factor indicated by color.}
\label{Extended Fig:Simulated mode}
\end{figure}

To achieve controllable lateral mode confinement, we adopt a heterogeneous structure in which the hole size in the barrier region is reduced to 0.95 times that in the core region, as shown in the inset of Fig.~\ref{Extended Fig:Simulated mode}b. The resulting band structures near the $\Gamma$ point are presented in Fig.~\ref{Extended Fig:Simulated mode}d,e. The resonant wavelength at the $\Gamma$ point in the barrier region is longer than that in the core region by about 26~nm, thereby forming a heterostructure that provides effective lateral feedback~\cite{Ge2019,Chen2022}. We further calculate the electric-field distributions of the full heterostructure for core-region unit-cell numbers $N_u=35$, 15, and 5 (Fig.~\ref{Extended Fig:Simulated mode}a-c).

\newpage
\clearpage
\subsection{Geometric phase and momentum-space polarization map}
\addsubtosi{Geometric phase and momentum-space polarization map}
\label{Note: Geometric phase}

\begin{figure}[htbp]
\refstepcounter{fig}
    \includegraphics[width=1\textwidth]{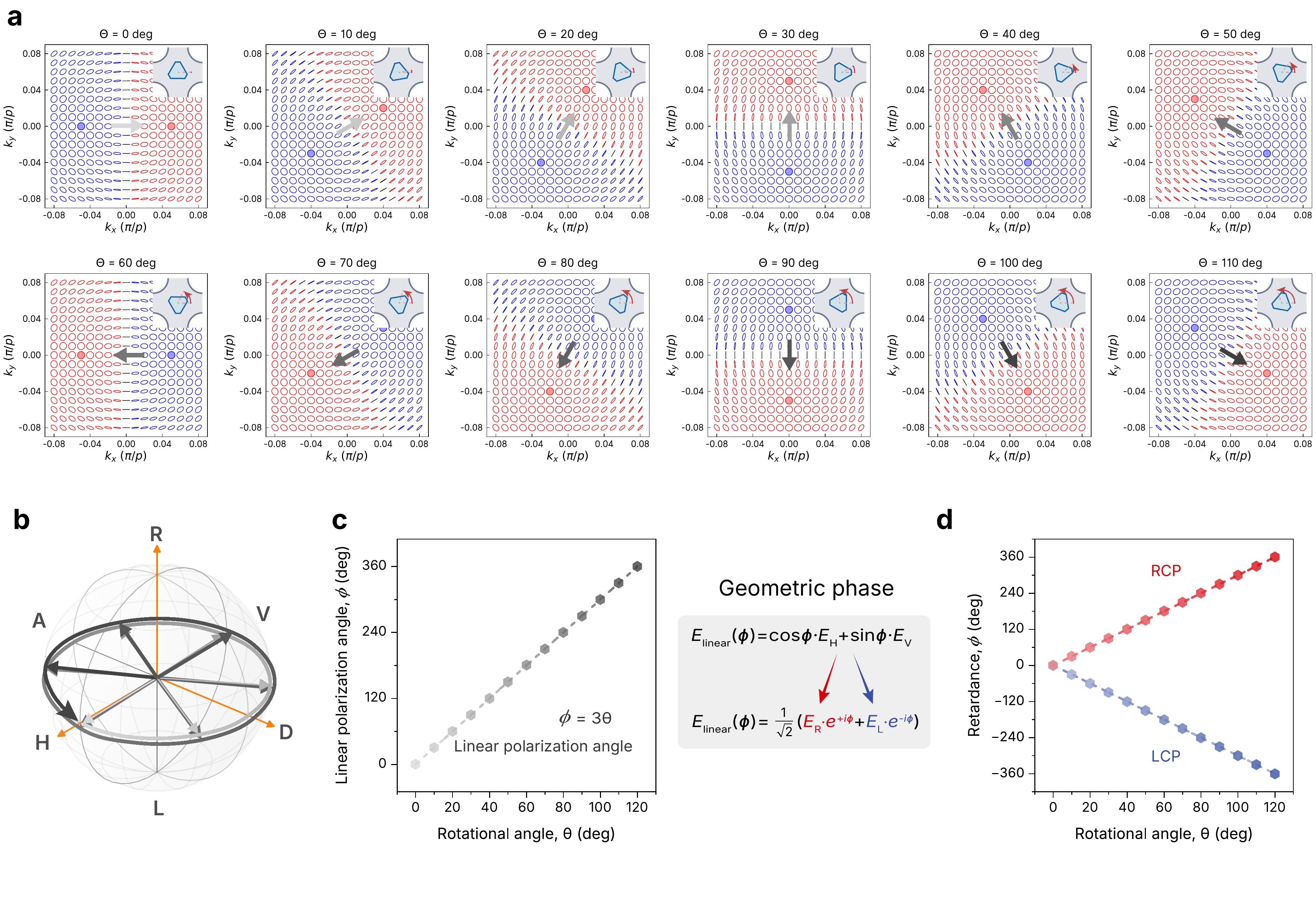} 
    \caption{\textbf{Eigen-polarization and geometric-phase analysis.}
    \textbf{a}, Evolution of the momentum-space eigen-polarization distributions for different geometric perturbation angles $\uptheta$. The black lines denote linear polarization. Red and blue ellipses denote right- and left-handed elliptical polarizations. LCP and RCP points are indicated by filled circles. The linear polarization angle at the $\Gamma$ point is marked by the gray arrow.
    \textbf{b}, Poincaré-sphere representation of the $\Gamma$-point eigen-polarization evolution. As $\uptheta$ increases from 0 to 120~deg, the eigen-polarization rotates along the equator of the Poincaré sphere by 720~deg.
    \textbf{c}, Linear polarization angle at the $\Gamma$ point as a function of $\uptheta$, which can be equivalently expressed in the circular basis and gives rise to the geometric phase shown in \textbf{d}.}
\label{Extended Fig:SFig_ pol_vectors}
\end{figure}

To extract the dependence of the geometric phase retardance on the perturbation rotation angle, we calculate the momentum-space polarization map. The truncated equilateral triangle reduces the in-plane symmetry of the unit cell from $C_4$ to $C_1$, thereby splitting the polarization singularity into a line of linear polarization and two circular polarization points (Fig.~\ref{Extended Fig:SFig_ pol_vectors}a). The gray arrow indicates the linear polarization angle at the $\Gamma$ point. As shown in Fig.~\ref{Extended Fig:SFig_ pol_vectors}b, when the perturbation rotation angle increases from 0 to 120~deg, the $\Gamma$-point eigenpolarization rotates by 720~deg around the equator of the Poincaré sphere. This originates from the in-plane $C_3$ symmetry of the truncated equilateral triangle. Figure~\ref{Extended Fig:SFig_ pol_vectors}c shows the extracted linear polarization angle as a function of the perturbation angle, revealing a linear relationship, $\phi=3\uptheta$. Accordingly, the linear polarization can be written as $E_{\text{linear}}(\phi)=\cos\phi\,E_{\mathrm{H}}+\sin\phi\,E_{\mathrm{V}}$. Expressed in the circular-polarization basis, where $E_{R}=\frac{1}{\sqrt{2}}(E_{H}-iE_{V})$ and $E_{L}=\frac{1}{\sqrt{2}}(E_{H}+iE_{V})$, this becomes $E_{\text{linear}}(\phi)=\frac{1}{\sqrt{2}}\left(E_{\mathrm{R}}e^{+i\phi}+E_{\mathrm{L}}e^{-i\phi}\right)$. This indicates that the evolution of the linear polarization at the $\Gamma$ point leads to opposite phase accumulation in the LCP and RCP components, giving rise to the geometric phase (Fig.~\ref{Extended Fig:SFig_ pol_vectors}d).

\newpage
\clearpage
\subsection{Spin-momentum-locked chiral emission}
\addsubtosi{Spin-momentum-locked chiral emission}
\label{Note: chiral emission}

To further demonstrate the flexibility of our platform, we also design and characterize spin-momentum-locked chiral emission from our quantum metasurface. Specifically, we implement a blazed-grating-like one-dimensional distribution of the perturbation rotation angle, defined as $\uptheta=n\cdot120/5$~deg, where $n$ is the unit-cell index. According to the geometric-phase relation, this phase gradient introduces opposite phase accumulations to the LCP and RCP components, thereby steering them into opposite momentum-space directions.

\begin{figure}[htbp]
\refstepcounter{fig}
    \includegraphics[width=1\textwidth]{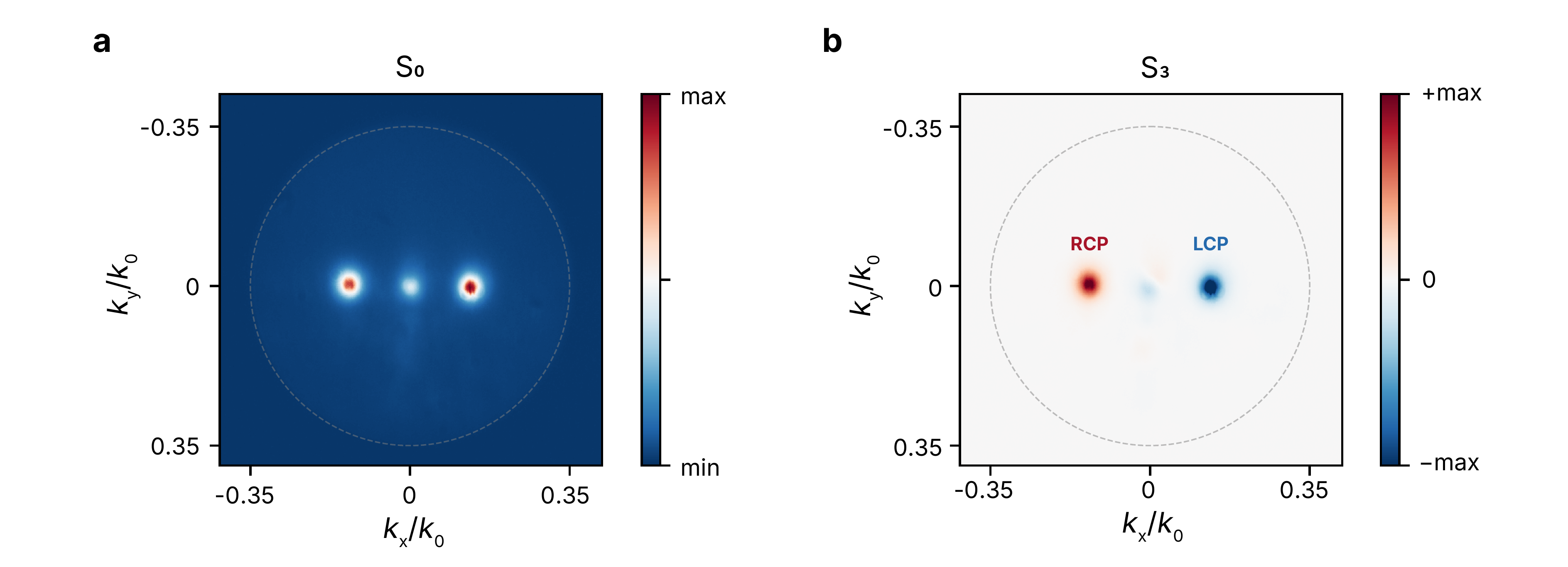} 
    \caption{\textbf{Momentum-space images of the band-edge-mode emission.}
    \textbf{a}, Measured normalized total intensity $S_0$.
    \textbf{b}, Measured circular-polarization Stokes parameter $S_3$. The gray dashed circle indicates the collection limit of the objective lens with a numerical aperture of 0.7.}
\label{Extended Fig:chiral exp}
\end{figure}

To quantify the spin selectivity of the emitted light, we evaluate the circular-polarization Stokes parameter $S_3=(I_{RCP}-I_{LCP})/(I_{RCP}+I_{LCP})$, where $I_{RCP}$ and $I_{LCP}$ denote the intensities of the right- and left-circularly polarized emission, respectively. Figure~\ref{Extended Fig:chiral exp}a shows the measured normalized total intensity $S_0$, while Fig.~\ref{Extended Fig:chiral exp}b shows the corresponding $S_3$ map in momentum space. The opposite signs of $S_3$ at opposite emission angles along the $k_x$ direction confirm the successful realization of spin-momentum-locked chiral steering.

\newpage
\clearpage
\subsection{Measurements and simulations of the unperturbed metasurface}
\addsubtosi{Measurements and simulations of the unperturbed metasurface}
\label{Note: unperturbed BIC}

\begin{figure}[htbp]
\refstepcounter{fig}
    \includegraphics[width=1\textwidth]{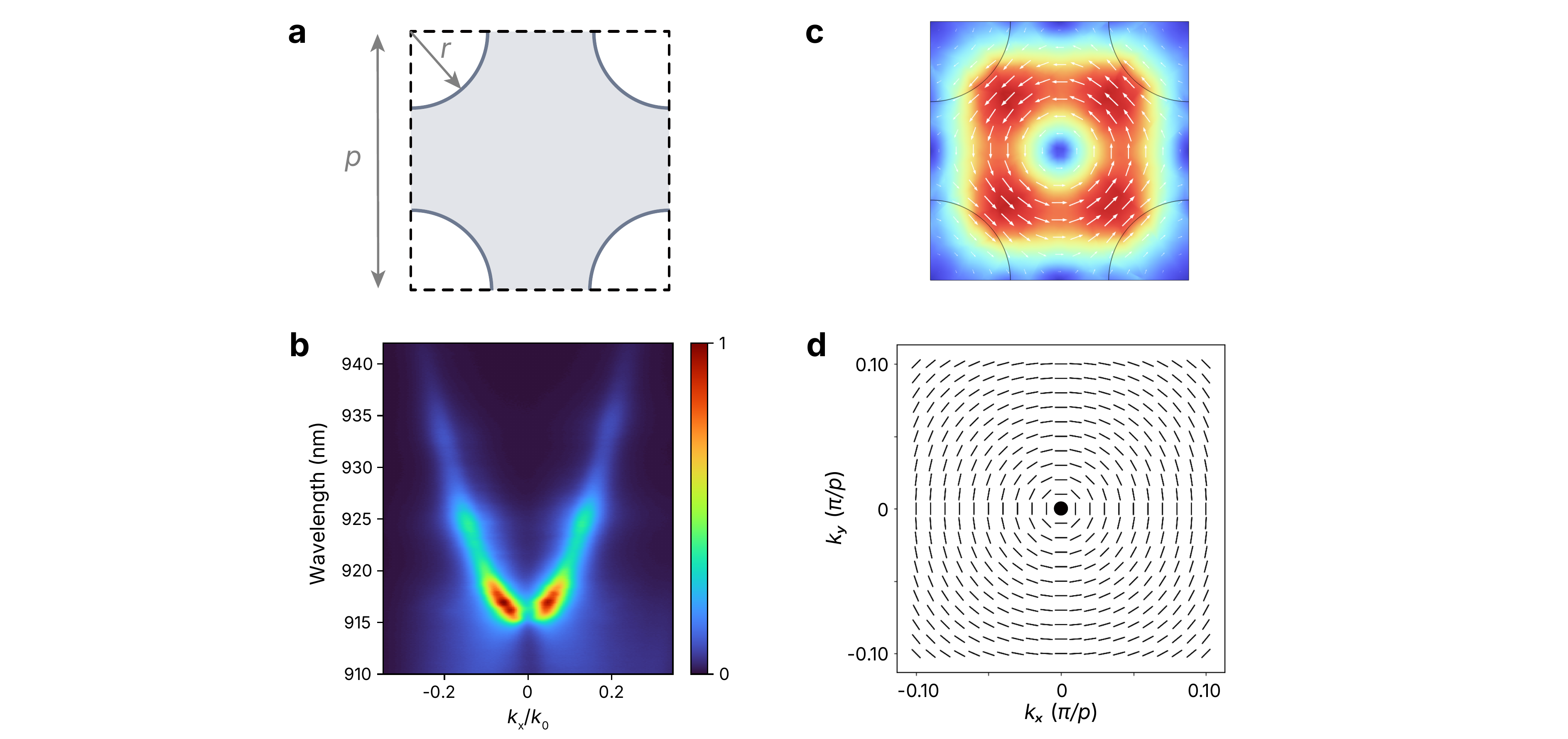} 
    \caption{\textbf{Measurements and simulations of the unperturbed metasurface.}
    \textbf{a}, Schematic illustration of the unit-cell geometry.
    \textbf{b}, Experimentally measured photonic band structure.
    \textbf{c}, Simulated electric-field distribution of the BIC.
    \textbf{d}, Simulated momentum-space polarization distribution.}
\label{Extended Fig:normBIC exp}
\end{figure}

For comparison, we also fabricate and measure the unperturbed metasurface supporting a symmetry-protected BIC (Fig.~\ref{Extended Fig:normBIC exp}a). The measured photonic band structure is shown in Fig.~\ref{Extended Fig:normBIC exp}b, where the emission is strongly suppressed at the $\Gamma$ point. This behaviour is a characteristic signature of the symmetry-protected BIC: although the mode exists at the band edge and above the light cone, it remains decoupled from the free-space radiation channel. We further simulate the mode profile and its momentum-space polarization distribution. As shown in Fig.~\ref{Extended Fig:normBIC exp}c, the electric-field distribution exhibits the characteristic $C_4$ symmetry. Meanwhile, the momentum-space polarization map in Fig.~\ref{Extended Fig:normBIC exp}d shows a polarization singularity at the $\Gamma$ point. 

\newpage
\clearpage
\subsection{Two-dimensional confocal raster-scan imaging}
\addsubtosi{Two-dimensional confocal raster-scan imaging}
\label{Note: confocal raster scan}

To ensure that only a single QD is excited within the metasource, we perform two-dimensional confocal raster-scan imaging under p-shell excitation. By scanning the sample relative to the focused excitation spot and recording the spectrally filtered PL intensity within a 0.4~nm spectral detection window at each position, we map the spatial distribution of the QD emission across the entire device. The measured intensity map confirms that only one bright QD is effectively addressed within the metasource.

\begin{figure}[htbp]
\refstepcounter{fig}
    \includegraphics[width=0.95\textwidth]{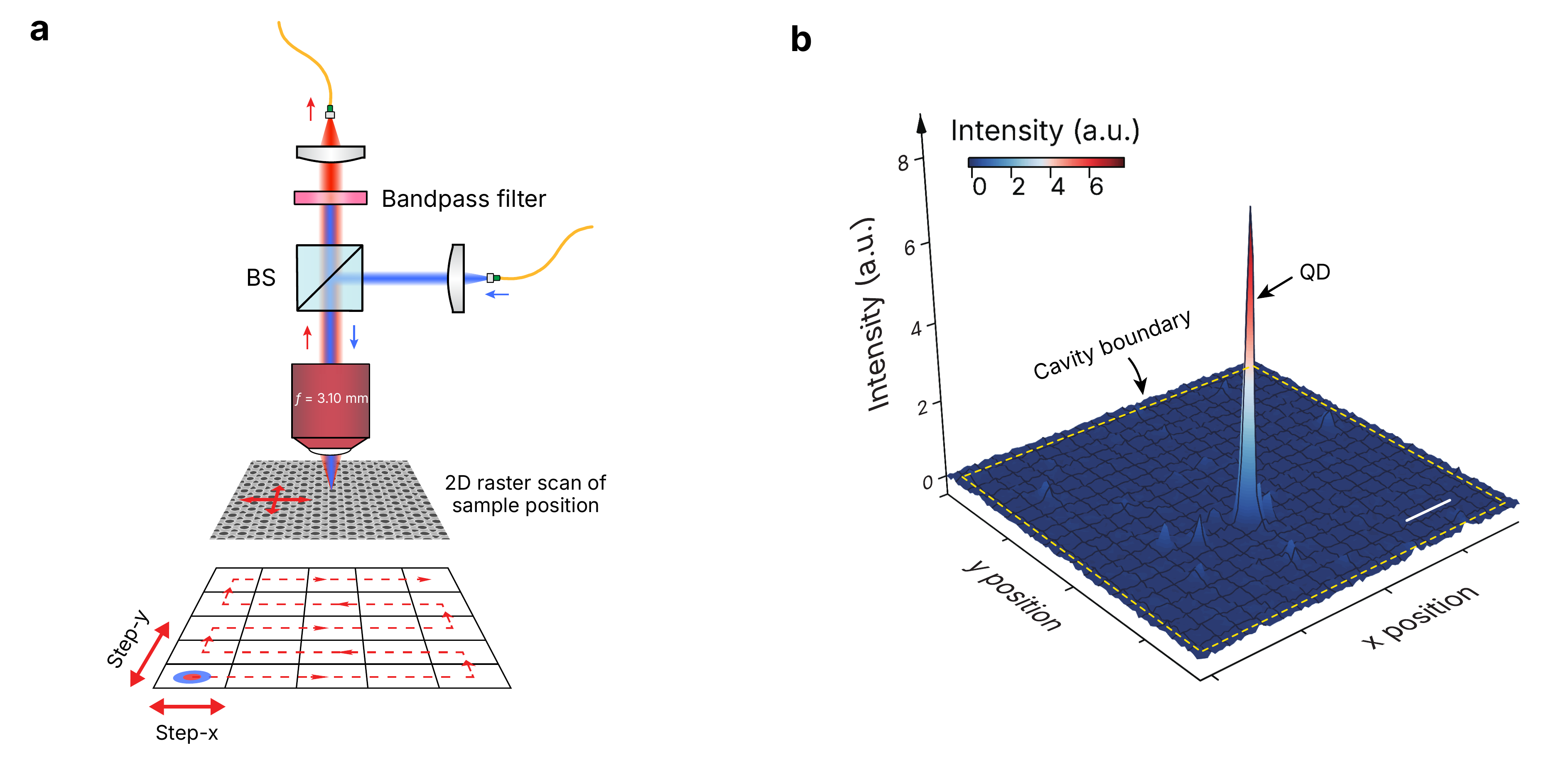} 

    \caption{\textbf{Two-dimensional confocal map of QD emission.}
    \textbf{a}, Schematic illustration of the confocal raster-scan measurement. Imaging is performed by scanning the metasurface sample across the focal plane of the objective lens along a serpentine path.
    \textbf{b}, Representative confocal image of QD emission obtained from the scan, showing that only a single QD is efficiently excited within the entire device.}
    
\label{Extended Fig:SFig_confocal_scan}
\end{figure}

\newpage
\clearpage
\putbib

\end{bibunit}

\end{document}